\documentclass[preprint,12pt]{elsarticle}

\usepackage{amssymb}

\usepackage[utf8]{inputenc} 
\usepackage[T1]{fontenc}
\usepackage{amsmath}
\usepackage{cleveref}
\usepackage{subcaption}
\usepackage{siunitx}

\journal{Computer Physics Communications}

\begin{document}

\begin{frontmatter}



\title{Verification of the Fourier-enhanced 3D finite element Poisson solver of
the gyrokinetic full-f code PICLS}

            
\author[inst1]{A.~Stier\corref{cor1}}
\ead{annika.stier@ipp.mpg.de}
\author[inst1]{A.~Bottino}
\author[inst2]{M.~Boesl}
\author[inst1]{M.~Campos Pinto}
\author[inst1]{T.~Hayward-Schneider}
\author[inst1]{D.~Coster}
\author[inst1]{A.~Bergmann}
\author[inst3]{M.~Murugappan}
\author[inst3]{S.~Brunner}
\author[inst3]{L.~Villard}
\author[inst1]{F.~Jenko}

\cortext[cor1]{Corresponding author}
\affiliation[inst1]{organization={Max-Planck Institute for Plasma Physics},
             addressline={Boltzmannstrasse~2},
             city={Garching},
             postcode={85748},
             state={Bavaria},
             country={Germany}}
\affiliation[inst2]{organization={Spark e-Fuels GmbH},
            addressline={Winterfeldtstrasse~18}, 
            city={Berlin},
            postcode={10781}, 
            state={Berlin},
            country={Germany}}
\affiliation[inst3]{organization={École Polytechnique Fédérale de Lausanne (EPFL) Swiss Plasma Center (SPC)},
            addressline={Rte Cantonale}, 
            city={Lausanne},
            postcode={CH-1015}, 
            state={State Two},
            country={Switzerland}}

\begin{abstract}
We introduce and derive the Fourier-enhanced 3D electrostatic field solver of the gyrokinetic full-f PIC code PICLS. The solver makes use of a Fourier representation in one periodic direction of the domain to make the solving of the system easily parallelizable and thus save run time. The presented solver is then verified using two different approaches of manufactured solutions. The test setup used for this effort is a pinch geometry with ITG-like electric potential, containing one non-periodic and two periodic directions, one of which will be discrete Fourier transformed. The results of these tests show that in all three dimensions the $L_2$-error decreases with a constant rate close to the ideal prediction, depending on the degree of the chosen basis functions.
\end{abstract}




\end{frontmatter}


\section{Introduction}
\label{introduction}
The PICLS (Particle-In-Cell Logical Sheath) code \cite{Boesl2020} is a full-f finite elements code with the purpose of simulating turbulence in the tokamak scrape-off layer. As indicated by its name, PICLS uses the PIC method with a gyrokinetic approximation for finite Larmor radius (FLR) effects of ions and drift-kinetic electrons.\\
So far, only 1D simulations conducted with PICLS were published in paper form \cite{Boesl2019a,Boesl2019b}. However, a field solver capable of handling 3D perturbations for electrostatic scenarios is already implemented and tested, with the caveat that this solver is operating on a linearized polarization equation. This is posing an exception to the otherwise full-f approach of PICLS. \\ In its 3D version, the solver is capable of operating in different geometries, using Cartesian  or polar coordinates.\\ 
The solver furthermore relies on at least one dimension of the problem being periodic and the finite element coefficients are discrete Fourier transformed (DFT). If the equilibrium density and the modulus of the background magnetic field are invariant along this dimension, e.g. the toroidal direction in a tokamak, the corresponding Fourier modes decouple from each other.
This approach drastically reduces the computational cost of the Poisson solver by replacing the full (3D dependent) solver problem to a number of independent smaller (2D dependent) matrix solver problems. The resulting algorithm is easy to parallelize and leads to an excellent scaling on present day supercomputers. Note that a similar approach has been successfully applied to $\delta f$ gyrokinetic simulations for closed field lines in tokamak  geometry using the ORB5 code \cite{Lanti2020}.
In the present paper we document and demonstrate this field solver of PICLS and determine its error with the method of manufactured solutions (MMS) \cite{Roache2001}. Out of the geometry options that PICLS currently provides for 3D perturbation runs (slab, helical slab, screw pinch and cylinder), we chose a screw pinch setup inspired by references \cite{Brunner1998} and \cite{Hatzky2002}, with an analytical potential field mimicking an ion temperature gradient (ITG) instability. To this end, \cref{sec:solver} lays out the derivation of the solver before the theoretical framework of the verification methods are described in \cref{sec:mms}. \Cref{sec:results} then details the findings gained from applying these methods.

\section{The 3D solver}\label{sec:solver}

The physical model of PICLS is based on the gyrokinetic particle Lagrangian
\begin{align} 
\label{eq:GKLagrangian}
        L=&&\sum_{s}\int \left(\left(\frac{e_s}{c}\Vec{A}+m_s v_{\parallel}\Vec{b}\right)\cdot\Dot{\Vec{R}}+\frac{m_s c}{e_s}\mu\Dot{\theta}-H_s \right)f_s \mathrm{d} W \mathrm{d} V \nonumber\\
        &&+ \int \frac{\tilde{E}^2-\tilde{B}_{\perp}^2}{8\pi} \mathrm{d} V 
\end{align}
written in CGS units, for particle species $s$ with charge $e_s$, mass $m_s$, parallel velocity $v_{\parallel}$, at gyrocenter position $\Vec{R}$ and gyro angle $\theta$ in a phase space composed of the velocity space $W$ and physical space $V$. $\Vec{A}$ is the  magnetic vector potential, $\Vec{b}$ the magnetic field unit vector, $\mu$ the magnetic moment, $f_s$ the distribution function, $\tilde{E}$ the perturbed electric field strength, $\tilde{B}$ the magnetic field strength and $c$ the speed of light. A general overview of the gyrokinetic theory and its application to plasma turbulence can be found in ref. \cite{ScottV2}. The full derivation of \cref{eq:GKLagrangian} is described in detail in ref. \cite{Tronko2017}. The physics content of \cref{eq:GKLagrangian} depends on the choice of the Hamiltonian $H_s$. Here we use the following Hamiltonian, corresponding to an electrostatic system in which electrostatic perturbations are assumed to have long perpendicular wavelengths as compared to the ion thermal Larmor radius \cite{Bottino2015},
\begin{align}
        H_s &=& H_{s,0} + H_{s,1} + H_{s,2}\nonumber\\
        H_{s,0} &=& \frac{m_s v_{\parallel}^2}{2} + \mu B\nonumber\\
        H_{s,1} &=& e_s J_{s,0} \Phi \nonumber\\
        H_{s,2} &=& - \frac{m_s c^2}{2B^2}|\nabla_{\perp}\Phi|^2
\end{align} 
where $J_{s,0}$ is the gyroaveraging operator and $\Phi$ the electrostatic potential.\\
PICLS imposes three approximations on the Lagrangian which do not affect self-consistency of the equations but cause limitations of the model.\\ Using the quasi-neutrality approximation, $E^2 \ll E_{E \times B}$, i.e. assuming that the energy associated to the magnetic perturbation is much smaller then the energy associated to the $E\times B$ motion, a linearized polarization approximation $f_s=f_{M,s}$ for $H_{s,2}$ and neglecting electromagnetic perturbations $\delta A_{\parallel}=0$, we arrive at a simplified particle Lagrangian
\begin{align}
\label{eq:simpleL}
        L =& \sum_{s} \int \left( \left( \frac{e_s}{c} \Vec{A} + m_{s} v_{\parallel} \Vec{b} \right) 
        \cdot \Dot{\Vec{R}} + \frac{m_{s} c}{e_{p}} \mu \Dot{\theta} -H_{s,0}-H_{s,1} \right)\nonumber\\ 
        &f_{s} \mathrm{d} W \mathrm{d} V + \sum_{s} \int \frac{m_{s} c^{2}}{2B^2} |\nabla_{\perp} \Phi|^{2} f_{M,s}  \mathrm{d} W \mathrm{d} V
\end{align}
in which now $\vec{A}$ and $\vec{B}$ refer to the background magnetic field only.
The equation governing the evolution of the electric field is constructed by setting the functional derivative of $L$ with respect
to $\Phi$ to zero in order to minimize the action integral (see Chapter 5.9 of ref. \cite{ScottV2}). This equation is
called polarization equation (or gyrokinetic Poisson equation) as it balances a polarisation density with the gyrocenter charge density.
However, the system described by \cref{eq:simpleL} is by definition charge-neutral, having left the term $\propto \int \tilde{E}^2 \mathrm{d} V $
out of the Lagrangian. In this specific case the polarisation equation is sometimes called {\it quasi-neutrality} equation.

The discretisation and solution of the polarisation equation is the subject of this work. Note that in the variational framework of the gyrokinetic theory,  the discretization can be combined with the variational principle. The most natural method to do this is given by the Galerkin approximation, which consists in doing the variations over functions constrained to remain in a finite-dimensional function space.
This naturally leads to a finite-element approximation of the fields \cite{Bottino2015}.
In the field solver stage of the PIC cycle, the electric field is calculated during the so-called charge assignment, in which the charge associated
to the particles is projected on the finite element basis, following a procedure described in ref. \cite{Bottino2015}. Although the knots of the basis splines are not theoretically constricted in their arrangement, we chose to use evenly spaced knots for the following investigation.\\ The goal of this solver is to obtain the electrostatic potential $\Phi$ formed by the charge distribution, to then be able to determine the electrostatic field using
\begin{equation}
\label{eq:E}
    E = -\nabla \Phi
\end{equation}
Thus, $\Phi$ needs to be known in order to calculate the forces acting on the particles at their respective positions so they can be moved correctly in the particle pusher stage of the PIC cycle. We want to remark that the particle push is always done in Cartesian coordinates, independent of the coordinate system chosen for the field solver. This choice is motivated by the final  goal of building a field line independent code, able to simulate plasmas even in the absence of properly defined field lines, e.g. at the separatrix of a tokamak.
Nevertheless, since we will explain and test the solver with a pinch setup, the following elaborations will be based on cylindrical coordinates.

\subsection{General Method}
\begin{figure}
    \centering
    \includegraphics[width=0.5\linewidth]{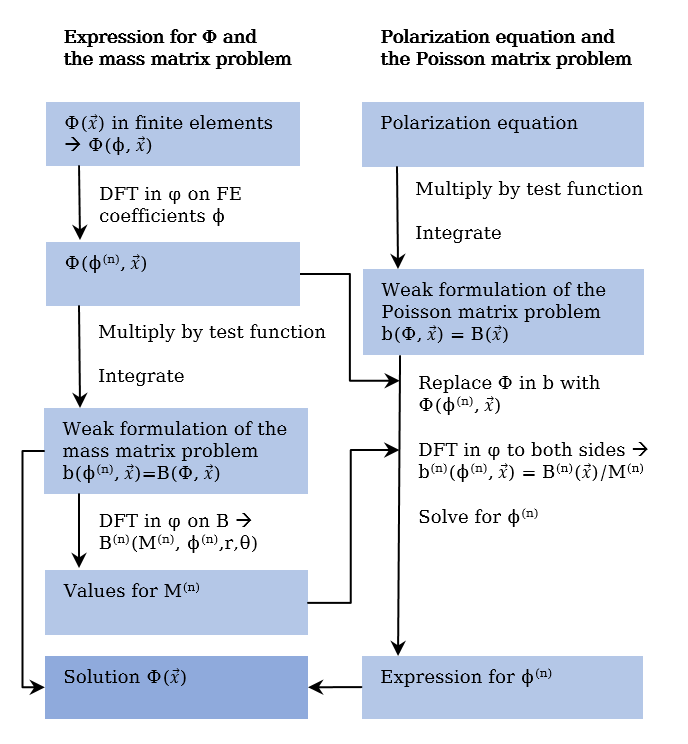}
    \caption{Steps towards the derivation of $\Phi$ in the solver.}
    \label{fig:flowchart}
\end{figure}

To get started, we will derive an expression for $\Phi$ in finite elements that expresses $\Phi$ through a series of Fourier coefficients $\phi^{(n)}$ of the initial finite element spline coefficients $\phi$. We use this expression as starting point to set up a weak formulation of the mass matrix problem.\\ Doing the same for the polarization equation and using the Fourier expression for $\Phi$ that was just derived, we can obtain a formulation that only contains $\phi^{(n)}$ and a quantity $M^{(n)}$ that can be determined from the mass matrix problem of the finite element $\Phi$ expression. Plugging in the quantity $M^{(n)}$ allows us to determine the Fourier coefficients $\phi^{(n)}$ of the spline coefficients $\phi$ of the potential. Knowing those, the weak formulation of the mass matrix problem provides the solution for $\Phi$. These steps are summarized in \cref{fig:flowchart} for general overview.

\subsection{Expression for $\Phi$ and the mass matrix problem}
\begin{figure}
    \centering
    \includegraphics[width=0.5\linewidth]{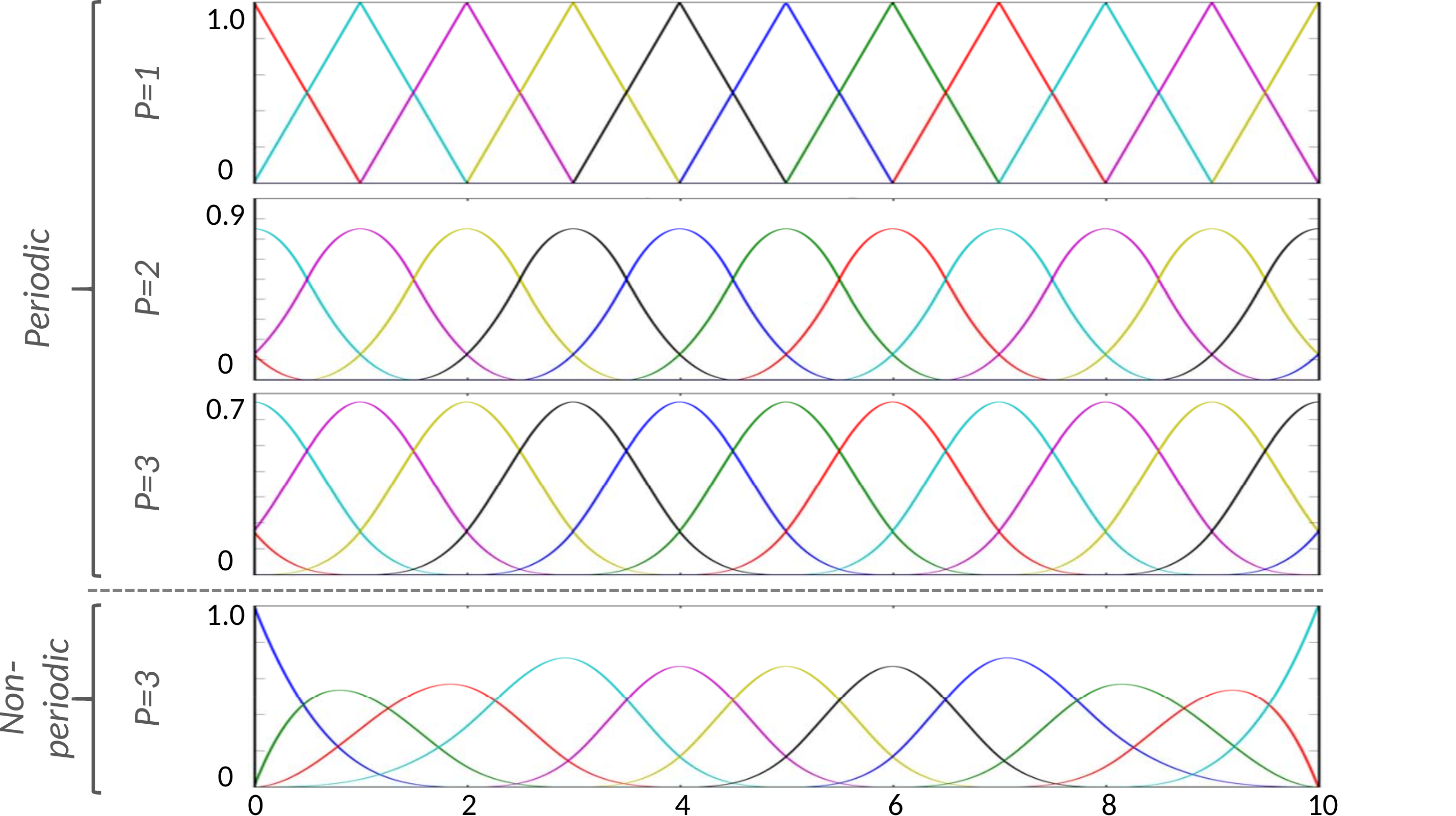}
    \caption{Periodic and non-periodic B-spline basis functions of different degrees $p$ on 10 intervals \cite{Boesl2020}.}
    \label{fig:splines}
\end{figure}

As mentioned above, the electrostatic potential in PICLS is discretized using finite elements (B-splines\cite{deBoor}). In this section we show how to compute the Fourier transformed B-splines coefficients, $\phi_{j,k}^{(n)}$, of the spline projection of some given arbitrary function $\Phi$.
Introducing the basis
functions of the finite-dimensional function space, $\tilde{\Lambda}_{\omega}$, all the functions in this space, including our electrostatic potential can be expressed as linear combinations of these basis functions
\begin{equation}
    \Phi(\Vec{x},t)=\sum_{\omega}^{n_{FE}}\phi_{\omega}(t)\tilde{\Lambda}_{\omega}(\Vec{x})
\end{equation}
with cylindrical coordinates $(r,\theta,\varphi)$. The basis functions $\tilde{\Lambda}_{\omega}$, can be written as the tensor product of 1D basis functions along $r, \theta$ and $\varphi$:
\begin{equation}
    \tilde{\Lambda}_{\omega}(\Vec{x})=\Lambda_{j}(r)\Lambda_{k}(\theta)\Lambda_{l}(\varphi)
\end{equation}
The basis functions in this case are B-splines of degree $p$. The symbols $j$, $k$ and $l$ are indexes ranging between 0 and the number of basis functions in their respective dimension $r$, $\theta$ and $\varphi$. Here, $r$ is the radial coordinate, $\theta$ the azimuthal coordinate and $\varphi$ is the axial coordinate of the cylindrical pinch, standing in for the toroidal dimension of a (linear) tokamak.\\
For periodic directions like $\theta$ and $\varphi$, the first and last spline of one dimension are identical and the number of splines is equal to the number of intervals in which the physical domain is partitioned. The periodic dimension $\varphi$ therefore has $n_{\varphi}+1$ knots and $n_{\varphi}$ intervals (and B-splines). For a non-periodic direction, such as $r$, the number of B-splines equals the number of intervals plus the spline degree $p$. The difference is illustrated in \cref{fig:splines}.

Furthermore, the equilibrium is axisymmetric, i.e. invariant in $\varphi$. Note that while the equilibrium is axisymmetric, the perturbation $\Phi$ is not constrained to be axisymmetric. Using this assumption for each time step $t$, the electrostatic potential is formulated as
\begin{equation} 
\label{eq:5.33}
    \Phi(r,\theta,\varphi,t)=\sum_{l=0}^{n_{\varphi}-1} \sum_{j=0}^{n_r+p-1} \sum_{k=0}^{n_{\theta}-1} \phi_{jkl}(t)\Lambda_{j}(r)\Lambda_{k}(\theta)\Lambda_{l}(\varphi)
\end{equation}
with sums over the number of different B-splines in each direction. Furthermore, the basis spline coefficients $\phi_{jkl}$ which constitute a discrete field on a grid can be Fourier-transformed in $\varphi$ which leads to the discrete Fourier transformed coefficients
\begin{equation} 
\label{eq:5.34}
    \phi_{jkl}(t)= \sum_{n=0}^{n_{\varphi}-1} \phi_{jk}^{(n)}(t)\exp{\left(\frac{2\pi i}{n_{\varphi}}nl\right)}
\end{equation}
with $n$ being the toroidal mode number.
Inserting this into \cref{eq:5.33} leads to
\begin{align}
    \label{eq:5.35}
    \Phi(r,\theta,\varphi,t) =& \sum_{l=0}^{n_{\varphi}-1}\sum_{j}\sum_{k}\sum_{n=0}^{n_{\varphi}-1}\phi_{jk}^{(n)}(t)\exp{\left(\frac{2\pi i}{n_{\varphi}}nl\right)} \nonumber\\ & \Lambda_{j}(r)\Lambda_{k}(\theta)\Lambda_{l}(\varphi)
\end{align}

Multiplying by a test function and integrating over the entire phase-space leads to the so called weak formulation of \cref{eq:5.35}.
\begin{align}
\label{eq:5.36}
        &\sum_{l=0}^{n_{\varphi}-1}\sum_{n=0}^{n_{\varphi}-1}\exp{\left(\frac{2\pi i}{n_{\varphi}}nl\right)}\int_{0}^{2\pi}\Lambda_{l'}(\varphi)\Lambda_{l}(\varphi)\mathrm{d}\varphi \nonumber\\ &\sum_{j}\sum_{k}\phi_{jk}^{(n)} \int \Lambda_{j'}(r)\Lambda_{k'}(\theta)\Lambda_{j}(r)\Lambda_{k}(\theta)J(r,\theta) \mathrm{d} r \mathrm{d}\theta \nonumber\\
        &=\int \Phi(r,\theta,\varphi) \Lambda_{l'}(\varphi)\Lambda_{j'}(r)\Lambda_{k'}(\theta) J(r,\theta) \mathrm{d} r \mathrm{d}\theta \mathrm{d}\varphi
\end{align}
Having used the B-spline tensor product $\tilde{\Lambda}_{\omega}(\Vec{x})$ as the test function with indexes $j'$, $k'$ and $l'$ for its components. Note that in \cref{eq:5.36} the right-hand side and left-side have been swapped as compared to \cref{eq:5.35}.
 Here $J(r,\theta)$ is the general expression of the coordinate Jacobian, which, in the specific case of polar coordinates, reads $J(r,\theta)=r$.
 This equation can be written in a compact form 
 \begin{equation}
B_{j'k'l'}=b_{j'k'l'}
\end{equation}
Using the definitions
\begin{equation}
b_{j'k'l'}=\int \Phi(r,\theta,\varphi) \Lambda_{l'}(\varphi)\Lambda_{j'}(r)\Lambda_{k'}(\theta) J(r,\theta) \mathrm{d} r \mathrm{d}\theta \mathrm{d}\varphi
\end{equation}
and
\begin{align}       
\label{eq:Bijk}
B_{j'k'l'}=&\sum_{l=0}^{n_{\varphi}-1}\sum_{n=0}^{n_{\varphi}-1}\exp{\left(\frac{2\pi i}{n_{\varphi}}nl\right)}\int_{0}^{2\pi}\Lambda_{l'}(\varphi)\Lambda_{l}(\varphi)d\varphi \nonumber\\ &\sum_{j}\sum_{k}\phi_{jk}^{(n)} \int \Lambda_{j'}(r)\Lambda_{k'}(\theta)\Lambda_{j}(r)\Lambda_{k}(\theta) r dr d\theta \nonumber\\
\end{align} 
 we can easily verify that \cref{eq:Bijk} has the form of of a discrete Fourier transform
\begin{equation}
\label{eq:5.40}    B_{j'k'l'}=\sum_{n=0}^{n_{\varphi}-1}B_{j'k'}^{(n)}\exp{\left(\frac{2\pi i}{n_{\varphi}}nl'\right)}
\end{equation}
with Fourier coefficients
\begin{align}
\label{eq:5.42}
    B_{j'k'}^{(n)} =& M^{(n)}\sum_{j}\sum_{k}\phi_{j'k'}^{(n)} C_{jk,j'k'}
\end{align}
and
\begin{align}
C_{jk,j'k'}=& \int \Lambda_{j'}(r) \Lambda_{k'}(\theta) \Lambda_{j}(r) \Lambda_{k}(\theta) r dr d\theta \nonumber\\  
\end{align}
$M^{(n)}$ are scalar coefficients for which, for each $l'$
\begin{align}
\label{eq:5.45}
    M^{(n)}\exp{\left(\frac{2\pi i}{n_{\varphi}}nl'\right)}=& \sum_{l=0}^{n_{\varphi}-1} \exp{\left(\frac{2\pi i}{n_{\varphi}}nl\right)} \int_{0}^{2\pi}\Lambda_{l'}(\varphi)\Lambda_l(\varphi)\mathrm{d}\varphi
\end{align}
The integral $C_{jk,j'k'}$ can be interpreted as the element of a matrix, usually called the {\it mass matrix}. The mass matrix is a square sparse matrix, whose rank depends on the number of splines and their degree. Therefore, \cref{eq:5.35} corresponds to a linear algebra problem in which a matrix equation needs to be solved. To solve the system for the Fourier coefficients $\phi_{j'k'}^{(n)}$, we still need to determine $M^{(n)}$ (see next section).\\
Nevertheless, if  the $M^{(n)}$ coefficients are known, \cref{eq:5.36} reduces to
\begin{equation}
\sum_{n=0}^{n_{\varphi}-1}B_{j'k'}^{(n)}\exp{\left(\frac{2\pi i}{n_{\varphi}}nl'\right)}=\sum_{n=0}^{n_{\varphi}-1}b_{j'k'}^{(n)}\exp{\left(\frac{2\pi i}{n_{\varphi}}nl'\right)}
\end{equation}
corresponding to a set of $n_\varphi$ independent Matrix equations, one for each Fourier mode
\begin{align}
  \sum_{j}\sum_{k} C_{jk,j'k'}\phi_{j'k'}^{(n)}=\frac{1}{M^{(n)}}b_{j'k'}^{(n)}
\end{align}
Therefore, the resulting set of equations can be straightforwardly parallelized by assigning different Fourier modes to different computational units.

\subsubsection{Calculating $M^{(n)}$}
We can shorten the expression for $M^{(n)}$, \cref{eq:5.45}, by exploiting the compact support of each B-spline
\begin{align}
   &M^{(n)}\exp{\left(\frac{2\pi i}{n_{\varphi}}nl'\right)}=\sum_{l=0}^{n_{\varphi}-1} \exp{\left(\frac{2\pi i}{n_{\varphi}}nl\right)} \int_{0}^{2\pi}\Lambda_{l'}(\varphi)\Lambda_l(\varphi)\mathrm{d}\varphi\nonumber\\
    =& \exp{\left(\frac{2\pi i}{n_{\varphi}}nl'\right)}\sum_{a=-p}^{p} \exp\left({\frac{2\pi i}{n_{\varphi}}na}\right) \int_{0}^{2\pi}\Lambda_{l'}(\varphi)\Lambda_{l'+a}(\varphi) \mathrm{d} {\varphi}
\end{align}
and using the symmetry of periodic splines, $\Lambda(x)=\Lambda(-x)$, and of the complex exponential
\begin{align}
   M^{(n)}=&2\sum_{a=1}^{p} \cos\left(\frac{2\pi}{n_\varphi} na\right) \int_{0}^{2\pi}\Lambda_{l'}(\varphi)\Lambda_{l'+a}(\varphi) \mathrm{d} {\varphi}
   \nonumber\\
   &+\int_{0}^{2\pi}\Lambda_{l'}(\varphi)\Lambda_{l'}(\varphi) \mathrm{d} {\varphi}
\end{align}
Applying a coordinate transformation $\varphi = (2\pi / n_{\varphi})\tau$ and introducing the notation
\begin{equation}
    m_a = \int_{0}^{n_{\varphi}} \Lambda_{l+a}(\tau)\Lambda_{l}(\tau)\mathrm{d}\tau     
\end{equation}
we obtain
\begin{equation}
\label{eq:Mn}
    M^{(n)}= \frac{2\pi}{n_\varphi} \left(2 \sum_{a=1}^{p} m_a\cos\left(\frac{2\pi}{n_\varphi} na\right) +m_0\right)     
\end{equation}
Using the B-spline formulation of polynomials of degree $p$ defined within a grid cell \cite{Hoellig2003}, 
$m_a$ can also be written as
\begin{equation}
    m_a = \sum_{j=0}^{p-a}\int_{0}^{1}  P_{j+a}^{(p)}(\tau)P_j^{(p)}(\tau)\mathrm{d}\tau
\end{equation}
This formulation can be easily evaluated to obtain the coefficients $m_a$ for a given B-spline type of degree $p$ and consequently compute the value of $M^{(n)}$, defined by \cref{eq:Mn}.
For the sake of completeness this is done for concrete values of $p$ in \cref{sec:appendixA}.

\subsection{Polarization equation and the Poisson matrix problem}
In an axisymmetric system, the gyrokinetic polarization equation for coordinates $(r,\theta,\varphi)$ and a single ion species is of the elliptical form
\begin{equation}
\label{eq:pola}
    -\nabla \cdot (N(r,\theta)\nabla_{\perp}\Phi) = \sum_{s=i,e} q_s n_{s}
\end{equation}
with
\begin{equation}
    N(r,\theta)=\frac{n_{M,s}(r)m_s c^2}{B(r,\theta)^2}
\end{equation}
and 
\begin{equation}
    \nabla_{\perp} = \nabla r \frac{\partial}{\partial r} + \nabla \theta \frac{\partial}{\partial \theta} 
\end{equation}

featuring the gyro-density $n_{s}$, the Maxwellian equilibrium gyrocenter density $n_{M,s}$, the particle mass $m_s$, speed of light $c$ and the particle charge $q_s$. The assumption of particle density $n_{M,s}(r)$  is a consequence of the initial assumption of $f_M$ in the Lagrangian and linearizes the equation at the cost of a compromise on the full-f approach. The benefit of this restriction is the convenient independence of $N$ from $\varphi$ in \cref{eq:poissonMform_l}. It will have to be revised in the future if the Maxwellian assumption of the particle density is to be abandoned.\\
As for the electrostatic potential in \cref{eq:5.35}, we set up the weak formulation of this problem and follow a similar approach as before, making use of integration by parts for the phase-space integral. However, the elliptic structure of the polarization equation implies that boundary conditions have to be applied in the non periodic directions. In the specific case considered in this paper, the radial direction $r$ is the only non periodic one and (zero) Dirichlet boundary conditions are assumed on both sides of the radial domain. Note that in general PICLS allows for up to two non periodic coordinates in which  Dirichlet (zero and nonzero) boundary conditions can be applied.
The discrete polarization equation is again a matrix equation of the form 
\begin{equation}
\label{eq:poissonMform}
B_{j'k'l'}=b_{j'k'l'}
\end{equation}
with the charge density
\begin{align}
\label{eq:poissonMform_r}
    b_{j'k'l'} =&\int \left(\sum_{s=e,i}q_s n_{s}(r,\theta,\varphi)\right)\nonumber\\ &\Lambda_{j'}(r)\Lambda_{k'}(\theta)\Lambda_{l'}(\varphi) r \mathrm{d} r \mathrm{d}\theta \mathrm{d}\varphi    
\end{align}
on the right hand side and the Poisson part 
\begin{align}
\label{eq:poissonMform_l}
    B_{j'k'l'} =& \int N(r,\theta)\nabla_{\perp}\Phi(r, \theta, \varphi)\nonumber\\ &\nabla_{\perp}(\Lambda_{j'}(r)\Lambda_{k'}(\theta))\Lambda_{l'}(\varphi) r \mathrm{d} r \mathrm{d} \theta \mathrm{d} \varphi
\end{align}
on the left hand side.\\
Using \cref{eq:5.35} to express $\Phi$ yields
\begin{align}
    B_{j'k'l'}=&& \sum_{l=0}^{n_{\varphi}-1}\sum_{n=0}^{n_{\varphi}-1}\exp{\left(\frac{2\pi i}{n_{\varphi}}nl\right)}\int_{0}^{2\pi}  \Lambda_{l'}(\varphi)\Lambda_{l}(\varphi)\mathrm{d}\varphi\nonumber\\ 
        &&\sum_{j'}^{n}\sum_{k'}^{n}\phi_{j'k'}^{(n)}(t) \int  N(r,\theta)\nabla_{\perp}(\Lambda_{j'}(r)\Lambda_{k'}(\theta))\cdot\nonumber\\
        &&\nabla_{\perp}(\Lambda_{j}(r)\Lambda_{k}(\theta))r \mathrm{d} r \mathrm{d} \theta
\end{align}

Following the same procedure described in the previous section the discrete problem reduces to 
\begin{equation}
\sum_{n=0}^{n_{\varphi}-1}B_{j'k'}^{(n)}\exp{\left(\frac{2\pi i}{n_{\varphi}}nl'\right)}=\sum_{n=0}^{n_{\varphi}-1}b_{j'k'}^{(n)}\exp{\left(\frac{2\pi i}{n_{\varphi}}nl'\right)}
\end{equation}
where now 
\begin{align}
    B_{j'k'}^{(n)} =& M^{(n)}\sum_{j}\sum_{k}\phi_{j'k'}^{(n)} A_{jk,j'k'}
\end{align}
with the mass matrix replaced by the stiffness matrix
\begin{align}
\label{eq:5.74}
    A_{j'k',jk}=&& \int N(r,\theta)\nabla_{\perp}(\Lambda_{j'}(r)\Lambda_{k'}(\theta))\nonumber\\
    &&\nabla_{\perp}(\Lambda_j(r)\Lambda_{k}(\theta)) r \mathrm{d} r \mathrm{d} \theta
\end{align}
corresponding to a set of $n_\varphi$ independent matrix equations, one for each Fourier mode
\begin{align}
\label{eq:finalsystem}
  \sum_{j}\sum_{k} A_{jk,j'k'}\phi_{j'k'}^{(n)}=\frac{1}{M^{(n)}}b_{j'k'}^{(n)}
\end{align}

It is possible to conveniently reuse the synoptic $M^{(n)}$ that was already calculated for the mass matrix problem, while $b_{j'k'}^{(n)}$ needs to be calculated at every time step by taking the discrete Fourier transform of the charge density vector, \cref{eq:poissonMform_r}.

\begin{figure}
    \centering
    \includegraphics[width=0.5\linewidth]{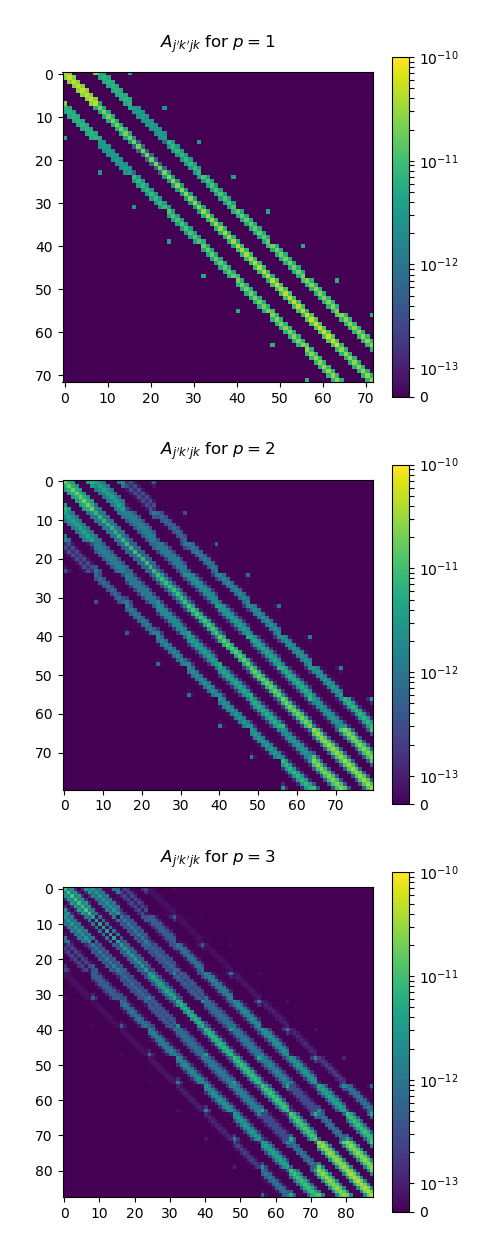}
    \caption{Structure of the matrix $A_{j'k',jk}$ for the present example of ITG instability in a screw pinch with a resolution of $n_r=n_{\theta}=8$ basis splines of differing degrees $p$.}
    \label{fig:matrices}
\end{figure}

The matrix elements $A_{j'k',jk}$, defined in \cref{eq:5.74}, are sparse and independent of $\varphi$ with a banded block structure of $2p+1$ block bands. For the two non-Fourier transformed dimensions $x$ and $y$, $A_{j'k',jk}$ is of rank $K_x \times K_y$, where $K_x=n_x$ if $x$ is periodic and  $K_x=n_x+p$ if $x$ is non periodic. Thus, in the case at hand, the rank of $A_{j'k',jk}$ is $(n_r + p)n_{\theta}$ as illustrated in \cref{fig:matrices}.\\
The right hand side of \cref{eq:pola}, given in its discrete form by \cref{eq:poissonMform_r}, contains the particle charge density and can be determined by performing the charge deposition step of the PIC cycle.
After that, the final system \cref{eq:finalsystem} of $n_{\varphi}$ matrix equations can be solved with the help of existing linear algebra packages to obtain values for the Fourier coefficients $\phi_{j'k'}^n$. In this paper, the linear algebra package \textit{Lapack} \cite{lapack99} has been used. Contrary to the full 3D problem in real space, our set of matrices is trivial to parallelize in the code implementation. This leads to a faster execution of the code for a problem of fixed resolution.
The potential $\Phi(r,\theta,\varphi)$ can finally be calculated by using expression (\ref{eq:5.35}).

\subsection{Poisson matrix in slab coordinates}

\begin{figure}
\centering
\hfill
\begin{subfigure}{0.45\linewidth}
    \includegraphics[width=\linewidth]{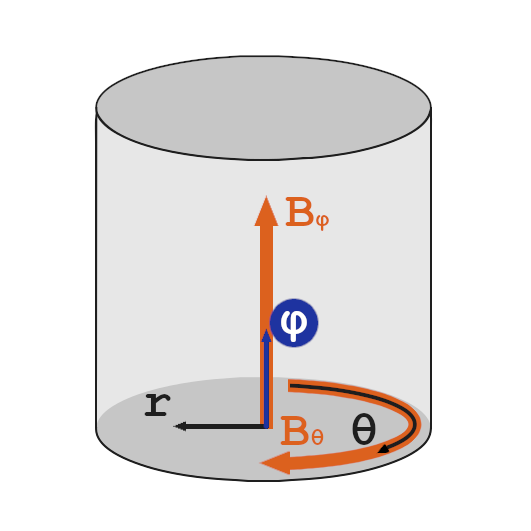}
    \caption{Pinch, periodic with respect to $\varphi$ and $\theta$; perpendicular gradient in $(r,\theta)$ plane.}
    \label{fig:pinch}
\end{subfigure}
\begin{subfigure}{0.45\linewidth}
    \includegraphics[width=\linewidth]{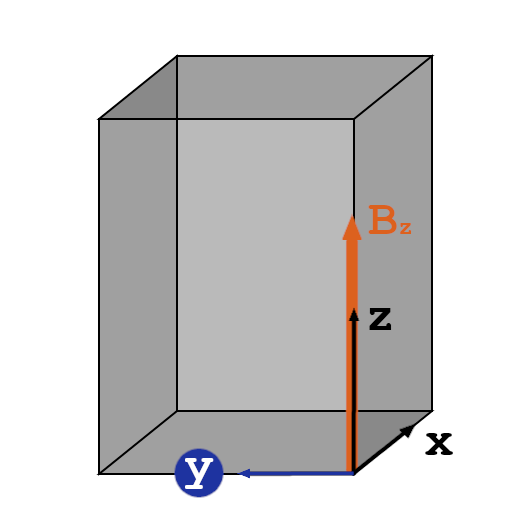}
    \caption{Slab, periodic with respect to $y$; perpendicular gradient in $(x,y)$ plane.}
    \label{fig:slab}
\end{subfigure}
\hfill
\caption{Depiction of the pinch and slab domain geometries available in PICLS and their allowed B-field components. The Fourier-transformed dimension of each is marked in blue and has to be periodic.}
\label{fig:geometries}
\end{figure}
Contrary to the cylindrical pinch geometry, where the magnetic field $B$ has a component in the periodic dimension $\varphi$, the slab and helical slab setups have the periodic dimension $y$ be part of the perpendicular gradient together with $x$. The $z$ coordinate is parallel to $B$ and non periodic. In general, the particles density (and consequently the polarization coefficient $N$) can vary along the field lines and the "radial" coordinate $x$, i.e. $N(x,z)$.
In the polarization equation (\ref{eq:pola}) we therefore use
\begin{equation}
    \nabla_{\perp} = \nabla x \frac{\partial}{\partial x} + \nabla y \frac{\partial}{\partial y}
\end{equation}
The expression for the left hand side $B_{j'k'l'}$, in which we now have partial derivatives also in the periodic dimension (assumed to be periodic n $[0,2\pi]$), is 
\begin{align}
        B_{j'k'l'}&=\sum_{l=0}^{n_y-1}\sum_{n=0}^{n_y-1}\exp{\left(\frac{2\pi i}{n_y}nl\right)}\int_{0}^{2\pi}\frac{\partial\Lambda_{l'}(y)}{\partial y}\frac{\partial\Lambda_{l}(y)}{\partial y}\mathrm{d} y \nonumber\\ & \sum_{j}\sum_{k}\phi_{jk}^{(n)}(t) \int N(x,z)\Lambda_{j}(x)\Lambda_{k}(z)\Lambda_{j'}(x)\Lambda_{k'}(z)\mathrm{d} x \mathrm{d} z\nonumber\\
        &+ \sum_{l=0}^{n_y-1}\sum_{n=0}^{n_y-1}\exp{\left(\frac{2\pi i}{n_y}nl\right)}\int_{0}^{2\pi}\Lambda_{l'}(y)\Lambda_{l}(y)\mathrm{d} y \nonumber\\ & \sum_{j}\sum_{k}\phi_{jk}^{(n)}(t) \int N(x,z)\frac{\partial\Lambda_{j}(x)}{\partial x}\Lambda_{k}(z)\frac{\partial\Lambda_{j'}(x)}{\partial x} \Lambda_{k'}(z)\mathrm{d} x \mathrm{d} z
\end{align}

Using the same notation of the pinch case, we can now define the two quantities:
\begin{align}
M^{(n)}=&&\sum_{a=-p}^{p} \exp\left({\frac{2\pi i}{n_{y}}na}\right) \int_{0}^{2\pi}\Lambda_{l'}(y)\Lambda_{l'+a}(y) \mathrm{d} y\
\end{align}
\begin{align}
\label{eq:Dn}
       D^{(n)}=&&  \sum_{a=-p}^{p} \exp{\left(\frac{2\pi i}{n_y}nl\right)} \int_{0}^{2\pi} \frac{\partial \Lambda_{l'+a}(y)}{\partial y}\frac{\partial \Lambda_{l'}(y)}{\partial y}dy
\end{align}
leading to
\begin{align}
    B_{j'k'}^{(n)} =& \sum_{j}\sum_{k} \left(D^{(n)} F_{jk,j'k'}+M^{(n)} G_{jk,j'k'}\right)\phi_{j'k'}^{(n)}
\end{align}
having define the two matrices 
\begin{align}
    F_{jk,j'k'}=&& \int N(x,z)\Lambda_{j'}(x)\Lambda_{k'}(x)\Lambda_j(x)\Lambda_{k}(z) \mathrm{d} x \mathrm{d} z
\end{align}
\begin{align}
    G_{jk,j'k'}=&& \int N(x,z)\frac{\partial\Lambda_{j}(x)}{\partial x}\Lambda_{k}(z)\frac{\partial\Lambda_{j}(x)}{\partial x} \Lambda_{k'}(z)\mathrm{d} x \mathrm{d} z
 \end{align}

corresponding once again to a set of $n_y$ independent matrix equations, one for each Fourier mode
\begin{align}
\label{eq:finalsystemslab}
  \sum_{j}\sum_{k} \left(D^{(n)} F_{jk,j'k'}+M^{(n)} G_{jk,j'k'}\right)\phi_{j'k'}^{(n)} = b_{j'k'}^{(n)}
\end{align}

\subsubsection{Calculating $D^{(n)}$}
Using the symmetry of the periodic splines' first derivative $\partial\Lambda(x)/\partial x=-\partial\Lambda(-x)/\partial x$ and of the complex exponential, \cref{eq:Dn} becomes 
\begin{align}
   D^{(n)}=&2\sum_{a=1}^{p} \cos\left(\frac{2\pi}{n_\varphi} na\right) \int_{0}^{2\pi}\frac{\partial\Lambda_{l'+a}(y)}{\partial y}\frac{\partial\Lambda_{l'}(y)}{\partial y}\mathrm{d} y
   \nonumber\\
   &+\int_{0}^{2\pi}\frac{\partial\Lambda_{l'}(y)}{\partial y}\frac{\partial\Lambda_{l'}(y)}{\partial y}\mathrm{d} y
\end{align}
Applying a coordinate transformation $y = (2\pi / n_y)\tau$ and introducing the notation
\begin{equation}
    d_a = \int_{0}^{n_y} \frac{\partial \Lambda_{l+a}(\tau)}{\partial \tau}\frac{\partial \Lambda_{l}(\tau)}{\partial \tau}\mathrm{d}\tau     
\end{equation}
we obtain
\begin{equation}
\label{eq:Dn2}
    D^{(n)}= \frac{n_y}{2\pi}\left(2\sum_{a=1}^{p} d_a\cos\left(\frac{2\pi}{n_\varphi} na\right)+d_0\right)     
\end{equation}
Using the B-spline formulation of polynomials of degree $p$ defined within a grid cell \cite{Hoellig2003}, $d_a$ can also be written as
\begin{equation}
    d_a = \sum_{j=0}^{p-a}\int_{0}^{1}\frac{\partial P_{j+a}^{(p)}}{\partial \tau}  \frac{\partial P_{j}^{(p)}}{\partial \tau}\mathrm{d}\tau
\end{equation}
It is important to notice that $D^{(0)}=0$ for any value of $p$. 
For the sake of completeness the actual values of  $D^{(n)}$ for $p=1,2,3$ are reported in \cref{sec:appendixB}.

\section{Method of Manufactured Solutions}\label{sec:mms}
The approach we use to verify our implemented solver is based on the Method of Manufactured Solutions (MMS) \cite{Roache2001,Oberkampf10}. To this end, a known or manufactured solution is provided e.g., by adding source terms to the equations. 
The deviation between the numerical solution of the equation and the provided manufactured solution is then checked.\\
In our specific case, this method is slightly adapted and applied to check whether the solver routines calculate the correct potential from a given charge distribution. As a test scenario we choose a screw pinch geometry with ITG-like perturbations.

\subsection{Testing the projection of the potential and the solver}
\label{sec:MassTest}
As elaborated in the previous section, PICLS solves the polarization equation in the matrix form \cref{eq:finalsystem} in order to obtain the Fourier transformed spline coefficients of the potential $\phi_{jk}^{(n)}$.\\ In case of MMS, those can be known and used to back-solve for $B_{jk}^{(n)}$ to provide the solver with a right hand side from which the normal solver routine can be started. We end up with a solved solution for $\Phi$ to compare with the manufactured analytical function that $B_{jk}^{(n)}$ was constructed from. In specific detail, the procedure follows these steps:
\begin{enumerate}
    \item We define an analytical potential, $\Hat{\Phi}=f_a$, for a specific test case. Here, we chose an ITG-instability-like scenario \cite{Brunner1998} which we split into three different functions, $f_a$ to be used separately for scans in $r$,$\theta$ and $\varphi$ respectively: 
    \begin{align}
        &f_1(r)=\tilde{a}(g(r)+\alpha r+\beta)\label{eq:mms1}\\
        &f_2(r,\theta)=\tilde{a}\sin(m_i\theta)(g(r)+\alpha r+\beta)\label{eq:mms2}\\
        &f_3(r, \varphi)=\tilde{a}\sin(n_i\varphi)(g(r)+\alpha r+\beta)\label{eq:mms3}
    \end{align}
    with
    \begin{align}
        &g(r)\equiv\exp\left(-\frac{1}{2}\left(\frac{r-r0}{\sigma_r}\right)^2\right)\nonumber
    \end{align}
    where $n_i=4$, $r_0=5$, $\sigma_r=1$ and $\tilde{a}=1$ are input parameters and
    $m_i=n_i q$ with $q$ being the screw-pinch safety factor. $\alpha$ and $\beta$ are constants which are determined by the boundary conditions in $r$. Dirichlet boundary conditions are used, which imply, taking for example \cref{eq:mms1},
    \begin{align}
        &\phi_i(0)=f_1(0)=0\nonumber\\
        &~~\rightarrow~~\beta=-g(0)\\
        &\phi_i(r_{\textrm{edge}})=f_1(r_{\textrm{edge}})=0\nonumber\\
        &~~\rightarrow~~\alpha=-\tilde{a}\frac{g(r_{\textrm{edge}})-g(0)}{r_{\textrm{edge}}}=0~~. 
    \end{align}
    The choice of the three  functions is motivated by the need of testing different aspects of the solver. The function $f_1$ allows for testing the behavior of the solver in the presence of non periodic boundary conditions, while $f_2$ is designed to test the quality of the spline projection with periodic splines. The function $f_3$ is a valid test for the discrete Fourier transform part of the solver and thus holds the most significance to prove the validity of our improved algorithm.    
    \item The analytic solution needs to be projected on a B-spline basis with the spline coefficients $\hat{\phi}_{jk}^{(n)}$ Fourier transformed in $\varphi$.
    The spline coefficients are calculated by solving the mass matrix problem \cref{eq:5.36}.
    \item Once the $\hat{\phi}_{jk}^{(n)}$ coefficients are known, a simple matrix multiplication provides the $\hat{B}_{jk}^{(n)}$ needed for the MMS tests:
      \begin{equation}
        \hat{b}_{jk}^{(n)}=M^{(n)}\sum_{j'k'} \hat{\phi}_{j'k'}^{(n)} A_{j'k',jk}.
      \end{equation}
    \item The coefficients $\hat{b}_{jk}^{(n)}$ are passed to PICLS (see \cref{eq:finalsystem}) and
      the normal solver routine is applied to calculate the corresponding potential spline coefficients ${\phi}_{j'k'}^{(n)}$,
      \begin{equation}
        \sum_{j'k'} {\phi}_{j'k'}^{(n)} A_{j'k',jk}= \frac{\hat{b}_{jk}^{(n)}}{M^{(n)}}~~.
      \end{equation}
      and per those the solution for ${\Phi}(r,\theta,\varphi)$ from \cref{eq:5.35}.
    \item $\Phi(r,\theta,\varphi)$ is evaluated and compared with the analytical input on a grid $(N_{r,\textrm{mms}},N_{\theta,\textrm{mms}},N_{\varphi,\textrm{mms}})$ of $(100/90/110)$ points , by defining the error as $L_2$ norm
      \begin{align}
      \label{eq:errors}
        L_2=&&\frac{\sqrt{\sum_{ijk}(\hat{\Phi}(r_i,\theta_j,\varphi_k)-{\Phi}(r_i,\theta_j,\varphi_k))^2}}{\sqrt{\sum_{ijk}\hat{\Phi}(r_i,\theta_j,\varphi_k)^2}}
      \end{align}
      with $i,j$ and $k$ being indexes ranging from 1 to $N_{r,\textrm{mms}},N_{\theta,\textrm{mms}}$ and $N_{\varphi,\textrm{mms}}$ respectively. We calculate $L_2$ norm for different values of $n_s$, $n_{\theta}$ and $n_{\varphi}$ and different spline orders $p$. 
\end{enumerate}

Due to the nature of this method the physical content of the Poisson matrix $A_{j'k',jk}$ is not included in the testing and the error values obtained from it are independent of the details  of $A_{j'k',jk}$. This means that so far, we have only evaluated the error related to the solver algorithm itself and to the projection $\Phi$ onto the spline basis. In the following we will refer to this kind of tests as {\it mass matrix based} test.

\subsection{Testing the physics of the Poisson matrix}
\label{sec:PoissTest}

In order to include the physics of the Poisson matrix into the verification, we need to construct the entire right hand side analytically. This is only possible for cases  in which the B-fields is twice differentiable and results in a test less general than the previous one. For the chosen screw pinch setup, the B-field conveniently meets those requirements. The analytical solution $\hat{\Phi}$ we want to obtain has to be a solution of the weak form of the continuous polarization equation \cref{eq:pola}. With normalizations $c=1$, $m_s=1$, a flat density profile $n_s=1$, and while using $J(r,\theta) = r$ and
\begin{align}
    \label{eq:nablanabla}
\nabla r\cdot\nabla r =1, \quad
\nabla \theta\cdot\nabla r =0, \quad
\nabla r \cdot\nabla \theta =0, \quad
\nabla \theta\cdot\nabla \theta =\frac{1}{r^2}
\end{align}
the weak formulation is 
\begin{align}
\label{eq:polaWeak}
        &&\int \frac{1}{B^2}\nabla_{\perp}\Phi(r, \theta,\varphi)\cdot \nabla_{\perp}\tilde{\Lambda}(r,\theta,\varphi) r\mathrm{d} r \mathrm{d} \varphi\nonumber\\
        =&& \int \rho(r,\theta)\tilde{\Lambda}(r,\theta,\varphi) r\mathrm{d} r \mathrm{d}\theta
\end{align}
equivalent to \cref{eq:poissonMform} with the abbreviation $\rho$ for the right hand side of the polarization equation. If an analytical solution for $\hat{\Phi}$ is known (e.g. $f_1$, $f_2$ or $f_3$ from \cref{sec:MassTest}), $\rho$ is 
\begin{equation}
    \label{eq:rho}
    \rho = -\frac{1}{r}\nabla_{\perp}\cdot\left(\frac{r}{B^2}\nabla_{\perp}\hat{\Phi} \right)
\end{equation}
Inserting this back into \cref{eq:polaWeak}, performing the integral by parts and assuming natural boundary conditions, we arrive at
\begin{align}
\label{eq:A3.9}
        &&\int \left(-\frac{1}{r}\nabla_{\perp}\cdot \left(\frac{r}{B^2}\nabla_{\perp}\Phi\right)\right)\tilde{\Lambda}r \mathrm{d} r \mathrm{d} \theta \nonumber\\
        =&& \int \frac{1}{B^2}\nabla_{\perp}\hat{\Phi}\cdot \nabla_{\perp}\tilde{\Lambda}r \mathrm{d} r \mathrm{d} \theta
\end{align}
Expanding \cref{eq:rho} yields
\begin{align}
\label{eq:MartinTestRho}
        \rho =&& -\frac{1}{r}\frac{1}{B^2}\frac{\partial \hat{\Phi}}{\partial r} + \frac{2}{B^3}\frac{\partial B}{\partial r}\frac{\partial \hat{\Phi}}{\partial r}\nonumber\\
        &&-\frac{1}{B^2}\frac{\partial^2\hat{\Phi}}{\partial r^2}-\frac{1}{B^2r^2}\frac{\partial^2 \hat{\Phi}}{\partial \theta^2}
\end{align}
The screw pinch B-field is independent of the poloidal angle $\theta$:
\begin{equation}
\label{eq:B}
    B=\sqrt{\left(\frac{r B_0}{R q_s(r)}\right)^2+B_0^2}
\end{equation}
In our tests, we assume $q_s(r)=q_s$ to be constant. This allows for 
\begin{equation}
    \frac{\partial B}{\partial r} = \frac{B_0}{B}\frac{r}{R^2q_s^2}
\end{equation}
We will use $\rho$ to derive coefficients $\hat{b}_{jk}^{(n)}$ that can be passed to the solver as in step 4 of \cref{sec:MassTest}. To mark the difference, we will call those coefficients $d_{jk}^{(n)}$ for this particular case.

In order to obtain $d_{jk}^{(n)}$ we do a charge assignment using  $\rho$
\begin{equation}
d_{j'k'l'}=\int \rho(r,\theta,\varphi) \Lambda_{l'}(\varphi)\Lambda_{j'}(r)\Lambda_{k'}(\theta) r \mathrm{d} r \mathrm{d}\theta \mathrm{d}\varphi
\end{equation}
followed by a discrete Fourier transform of it, being
\begin{equation}
d_{j'k'l'}=\sum_{n=0}^{n_{\varphi}-1}d_{j'k'}^{(n)}\exp{\left(\frac{2\pi i}{n_{\varphi}}nl'\right)}
\end{equation}

 $\hat{d}_{jk}^{(n)}$ can now be used as right hand side for the solver analogous to $\hat{b}_{jk}^{(n)}$ in step 4 of \cref{sec:MassTest} before. We can derive manufactured solutions for this extended test by plugging the analytical functions $f_1$, $f_2$ and $f_3$ as $\Phi$ into \cref{eq:MartinTestRho}. Note that in case of $f_3$ the expression varies with mode number $n$ due to the $\varphi$-dependence.  Here we present the results of a Poisson inclusive test for $f_2$ only.

\section{Results}\label{sec:results}
All of the tests for this publication were conducted on the high performance cluster \textit{RAVEN} featuring \textit{Intel Xeon IceLake-SP (Platinum 8360Y)} processors. 
The number of splines in the scanned direction was doubled until a clear deviation from the expected power law dependence of the $L_2$ norm could be noticed.\\ 
PICLS operates on 64-bit reals (52 bit mantissa) which corresponds to a precision of $10^{16}$ and sets the upper limit of accuracy for the following examination. However, through the sequence of arithmetic operations applied to these variables it is to be expected that the round-off error accumulates to higher levels.

\subsection{Scan in r with analytic potential $f_1$} \label{sec:f1_result}
\begin{figure}
    \centering
    \includegraphics[width=0.5\linewidth]{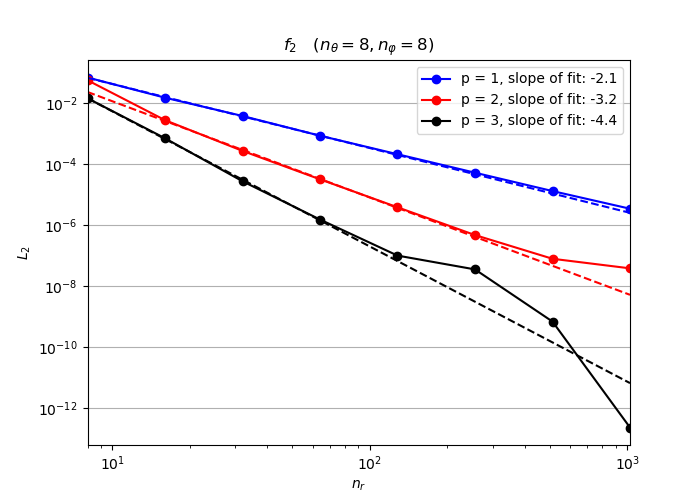}
    \caption{$L_2$-error for varying numbers of splines in $r$-direction and varying spline degrees $k$ using $f_1$ as the analytic reference.}
    \label{fig:f1_l2}
\end{figure}
\begin{figure}
    \centering
    \includegraphics[width=0.5\linewidth]{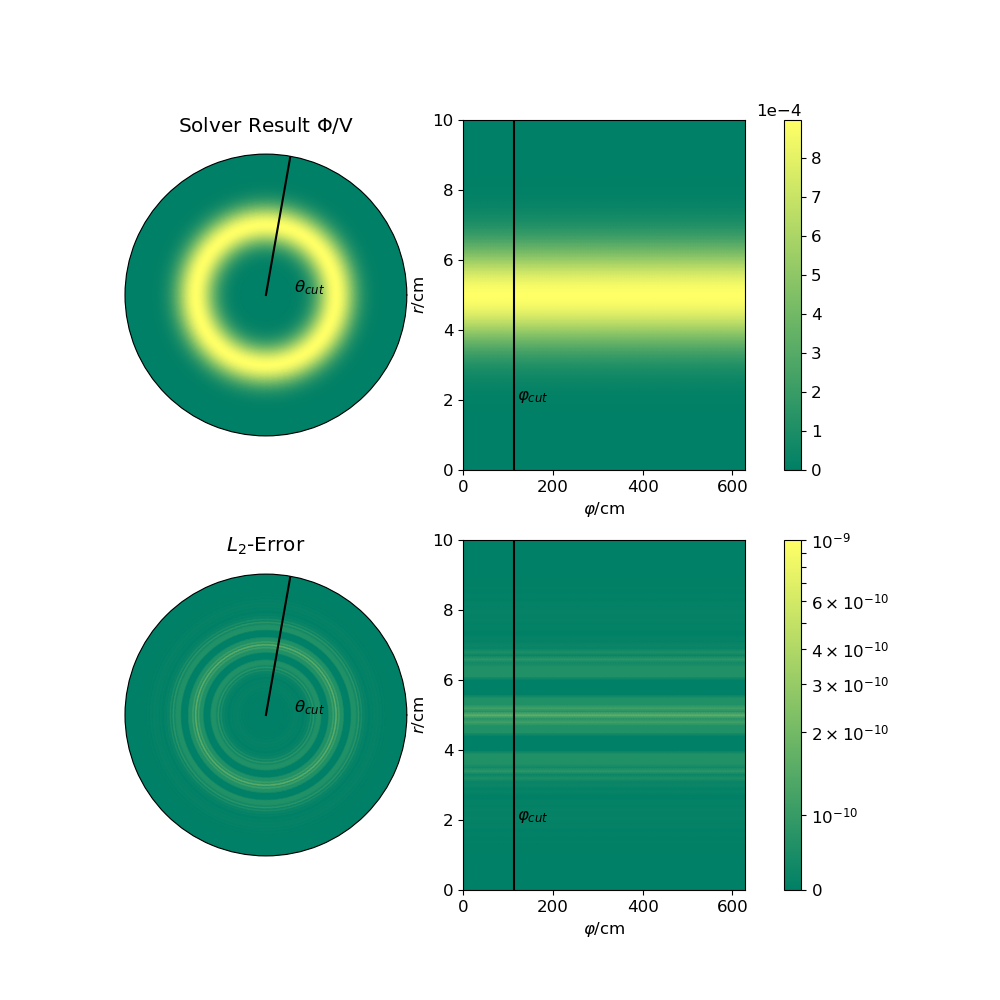}
    \caption{Solver result for $\Phi$ and the corresponding local contribution, $\epsilon$,  to the $L_2$-error for a resolution of $128\times 8 \times8$ intervals using $p=2$ and $f_1$ as the analytic reference. Cut at positions $\varphi=113.15$ cm and $\theta\approx 0.44\pi$.}
    \label{fig:candy_f1_l2}
\end{figure}

For the first mass matrix based MMS test (\cref{sec:MassTest}), the number of intervals in the periodic direction $\theta$ and $\varphi$ are kept constant while the $r$-resolution is doubled in each run. In order to get a significant range of $n_r$-values within the limits of available memory, $n_{\theta}$ and $n_{\varphi}$ were fixed at the low value of $8$. The minimum value for $n_r$ was also set to $8$ since for splines of degree $p=3$, the maximum spline degree in these tests, $8$ is the minimum number for which the matrix $A_{j'k',jk}$ of \cref{eq:5.74} is not fully occupied.\\
Up to $n_{r}=128$, the convergence of the error in number of splines follows a power law for all three spline degrees, with exponents $\approx$ \num{-2.1} for $p=1$, $\approx$ \num{-3.2} for $p=2$ and $\approx$ \num{-4.4} for $p=3$.
The order of convergence is expected to be equal to $p+1$ \cite{Li2010}  which matches the results shown here in rough approximation. From \cref{fig:f1_l2} it can be observed that the error deviates from the power law at the same value for every order of spline, as expected. An example of comparison between the analytical and
the calculated solution is shown if \cref{fig:candy_f1_l2} where the poloidal (top-left) and toroidal (top-right) cross sections of the calculated potential are plotted, together with the poloidal (bottom-left) and toroidal (botton-right) cross sections of the local  contribution to the $L_2$ error, $\epsilon=(\hat{\Phi}(r_i,\theta_j,\varphi_k)-{\Phi}(r_i,\theta_j,\varphi_k))^2$.

\subsection{Scan in $\theta$ with analytic potential $f_2$} \label{sec:f2_result}
\begin{figure}
\centering
\begin{subfigure}{0.45\linewidth}
    \includegraphics[width=\linewidth]{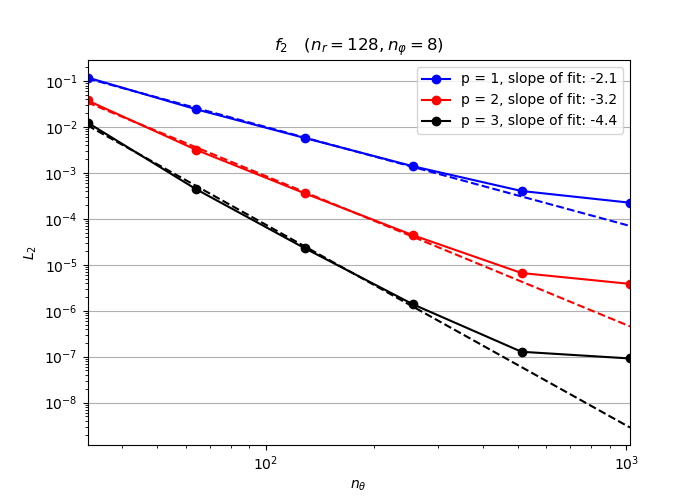}
    \caption{Mass matrix test.}
    \label{fig:mms_f2_l2}
\end{subfigure}
\hfill
\begin{subfigure}{0.45\linewidth}
    \includegraphics[width=\linewidth]{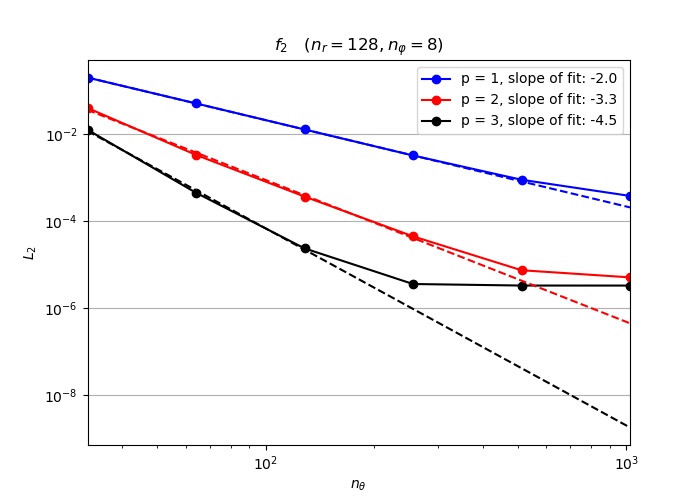}
    \caption{Poisson matrix test.}
    \label{fig:mms_f2_l2_Martin}
\end{subfigure}
\caption{$L_2$-error for varying numbers of splines in $\theta$-direction and varying spline degrees $k$ using $f_2$ as the analytic reference.}
\label{fig:mms_f2}
\end{figure}

\begin{figure}
\centering
\begin{subfigure}{0.45\linewidth}
    \includegraphics[width=\linewidth]{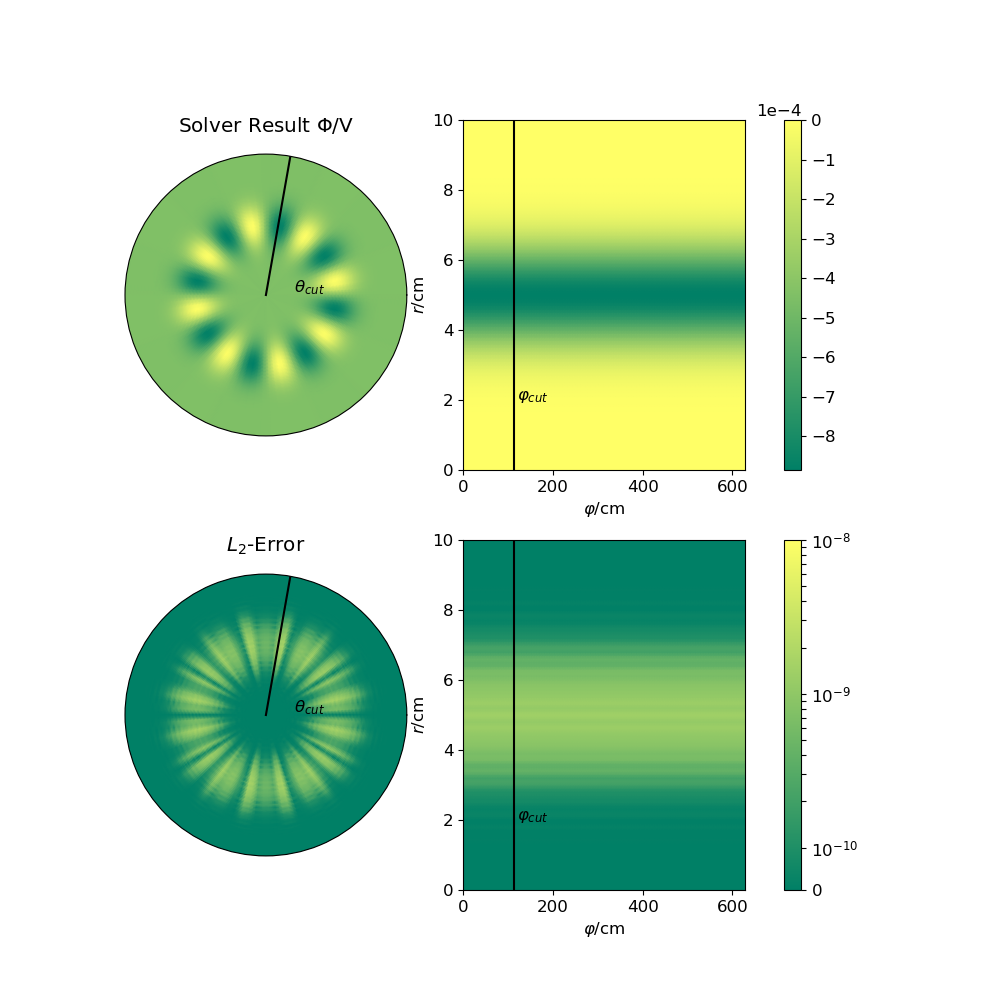}
    \caption{Mass matrix test.}
    \label{fig:candy_f2_l2}
\end{subfigure}
\hfill
\begin{subfigure}{0.45\linewidth}
    \includegraphics[width=\textwidth]{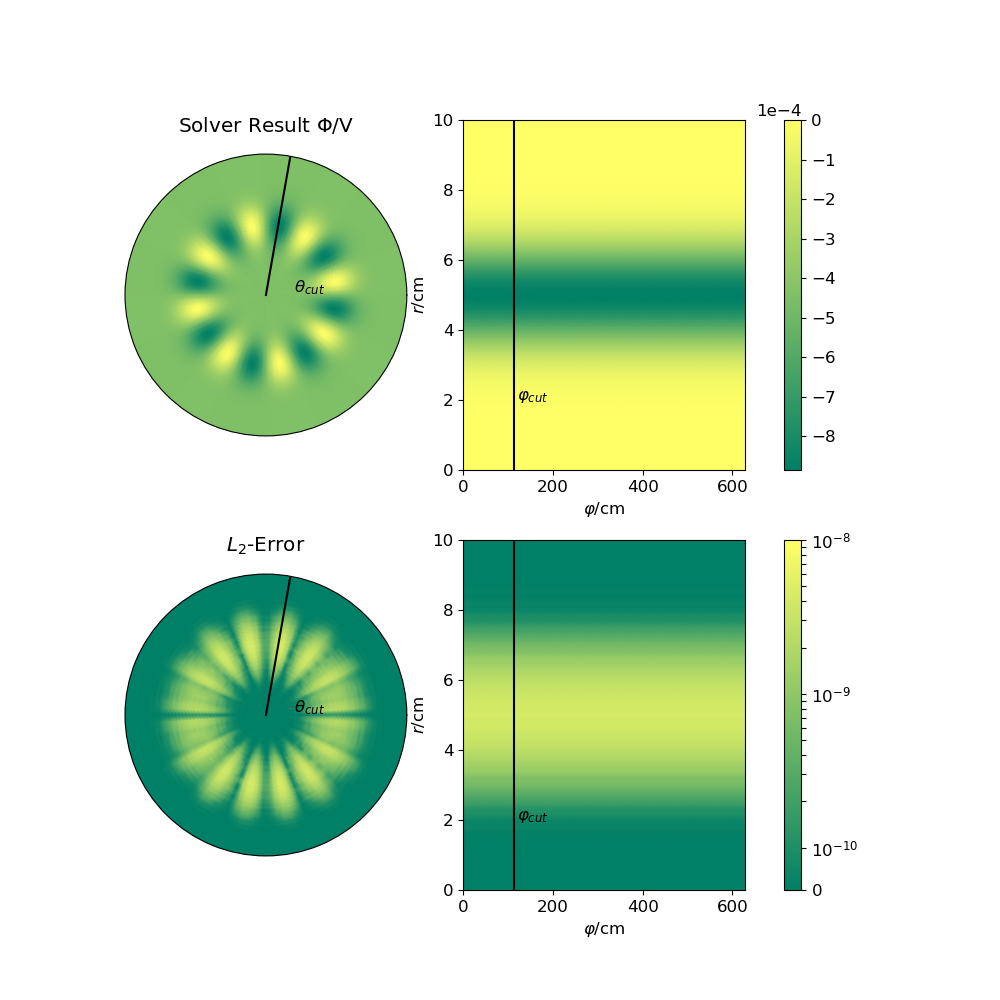}
    \caption{Poisson matrix test.}
    \label{fig:candy_f2_l2_Martin}
\end{subfigure}
\caption{Solver result for $\Phi$ and the corresponding local contribution, $\epsilon$, to the $L_2$-error for a resolution of $128\times 256 \times8$
using $p=2$ and $f_2$ as the analytic reference. Cut at positions $\varphi=113.15$ cm and $\theta\approx 0.44\pi$.}
\label{fig:candy_f2}
\end{figure}

For this mass matrix based test (\cref{sec:MassTest}), a scan in the number of periodic splines in $\theta$ is performed. $n_\varphi$ is set to the memory saving default value of $8$. Since $f_2$ has the same dependency in $r$ as $f_1$, the $r$-resolution  is kept fixed at $n_r=128$ with the intention of preventing a domination of the $\theta$-scan error by $n_r$-error contributions. Nevertheless, $L_2$ of $n_r=128$ contributes to the values we will obtain in this test and a difference in $p$ of $n_{\theta}$ for fixed $n_r$ can be expected.\\ The minimum of the range of $n_{\theta}$-values is determined by the mode number $m_i=n_i q = 16$ which has to be multiplied by $2$ to satisfy the Nyquist–Shannon criterion.\\
It becomes apparent from \cref{fig:mms_f2_l2} that the power law decrease of the error with number of splines is satisfied up to $n_\theta\approx 256$, above which it starts to deviate and saturates at a finite value determined by the chosen $n_r=128$ resolution. The measured exponents are \num{-2.1} for $p=1$, \num{-3.2} for $p=2$ and \num{-4.4} for $p=3$.\\
For the $f_2$ function, we have also performed a Poisson inclusive MMS test (\cref{sec:PoissTest}), by setting $\hat{\Phi} = f_2$ in \cref{eq:MartinTestRho}.
We conduct the same sweep in $n_{\theta}$ as for the mass matrix solver test. Involving the physics of the Poisson matrix leads in general to a higher $L_2$ error for all three spline degrees. Nevertheless, the expected power laws are recovered, with \num{-2.2} for $p=1$, \num{-3.3} for $p=2$ and \num{-4.5} for $p=3$.
The deviation from the power law no longer occurs at a roughly similar $n_{\theta}$ value but visibly shifts from higher $n_{\theta}$ for $p=1$ to lower $n_{\theta}$ for $p=3$, hinting that the inclusion of physics lessens the accuracy benefit caused by higher spline degrees.\\ Still, both investigations in \cref{fig:mms_f2} have in common, that the error seems to saturate at the same absolute value for all splines.~\Cref{fig:candy_f2} shows a comparison of relevant quantities for the MMS test described in this section.\\


\subsection{Scan in $\varphi$ with analytic potential $f_3$} \label{sec:f3_result}
\begin{figure}
    \centering
    \includegraphics[width=0.5\linewidth]{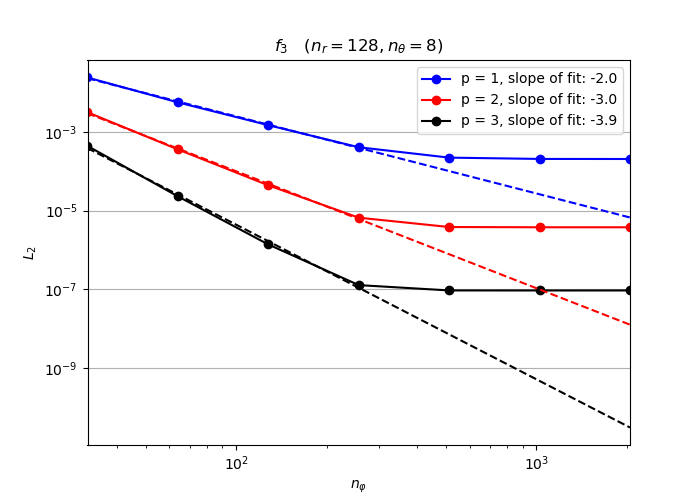}
    \caption{$L_2$-error for varying numbers of splines in $\varphi$-direction and varying spline degrees $k$ using $f_3$ as the analytic reference.}
    \label{fig:mms_f3_l2}
\end{figure}

\begin{figure}
    \centering
    \includegraphics[width=0.5\linewidth]{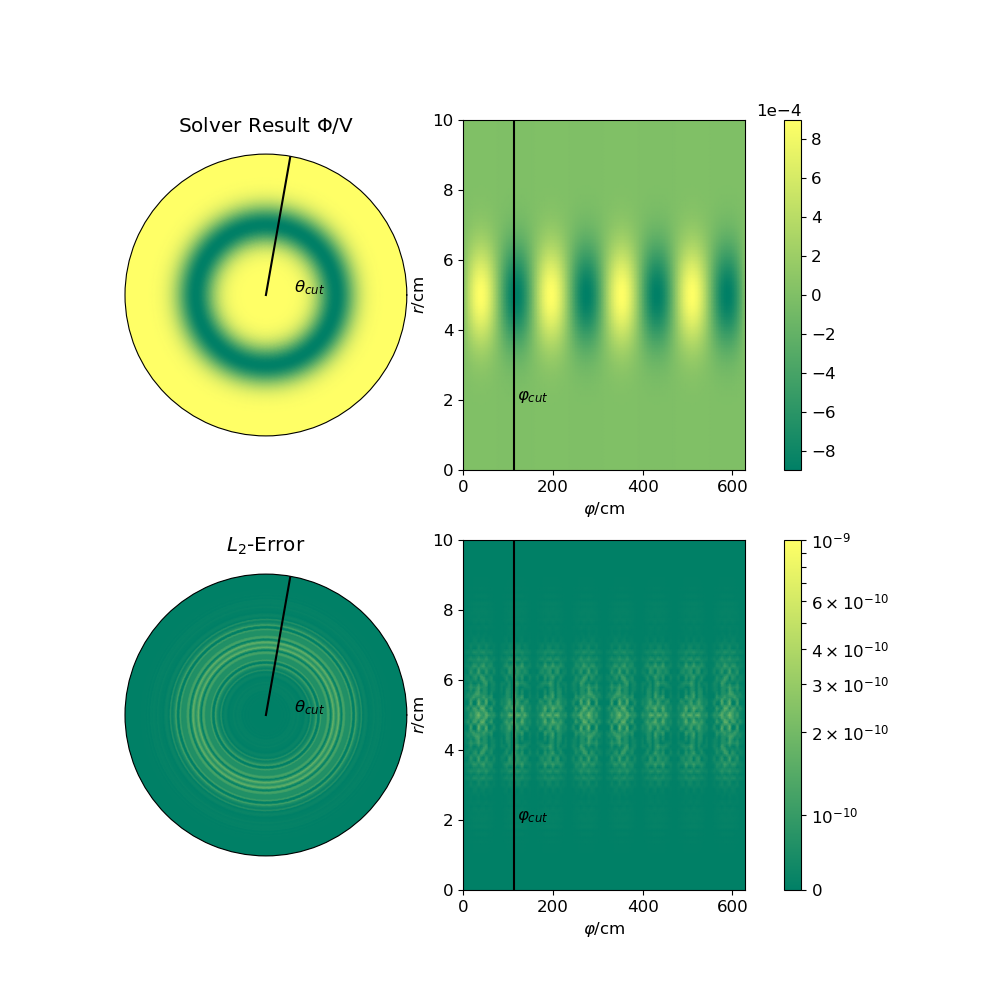}
    \caption{Solver result for $\Phi$ and the corresponding local contribution, $\epsilon$, to the $L_2$-error for a resolution of $128\times 8\times 256$ intervals using $p=2$ and $f_3$ as the analytic reference. Cut at positions $\varphi=114.2$ cm and $\theta\approx 0.44\pi$.}
    \label{fig:candy_f3_l2}
\end{figure}

Similar to $f_2$, the $f_3$ based mass matrix test has a dependence in $r$ and $n_r$ has to be kept at $128$.\\
To choose the starting point of the $n_{\varphi}$-scan we applied the Nyquist-Shannon criterion as for $f_2$ in \cref{sec:f2_result}, now for a mode number of $n_i=8$, and additionally took into account the applied Fourier transformation by going to the next higher exponent of $2$ which led to a minimum $n_{\varphi}$ of $32$.\\
\cref{fig:mms_f3_l2} shows similar trend as \cref{fig:mms_f2_l2}, despite the fact that now the Fourier transform based part of the solver is used.
The power laws are once again consistent with the theory, having exponents \num{-2.0} for $p=1$, \num{-3.0} for $p=2$ and \num{-3.9} for $p=3$.
The error saturation comes into effect at $n_{\varphi}=256$ similar to the $f_2$ case and it is related to the choice of $n_r=128$. The poloidal and toroidal cross sections
of the computed solution and of $\epsilon=(\hat{\Phi}(r_i,\theta_j,\varphi_k)-{\Phi}(r_i,\theta_j,\varphi_k))^2$ for the case $n_\varphi=256$ are shown in \cref{fig:candy_f3_l2}.

\section{Conclusion}\label{sec:conclusion}
In this work we have presented the Fourier-enhanced 3D B-spline based field solver implemented in the the gyrokinetic full-f PIC code PICLS. This algorithm is based on the observation that the mass matrix and any differential operator along any direction invariant by translation, e.g. the toroidal direction in a cylindrical pinch or in a tokamak, is effectively a convolution  of the corresponding finite element indices (assuming an equidistant grid). Given that the (discrete) Fourier transform of a convolution is the product of the Fourier transforms, these operators in Fourier representation become purely multiplicative, i.e. are represented by diagonal matrices. 
We showed by means of the method of manufactured solutions that the $L_2$ error norm of the Fourier-enhanced 3D finite element Poisson solver of PICLS diminishes for improving resolution with a rate close to the ideal one of $p+1$. This observation can be equally made for non-periodic, periodic, and Fourier-transformed periodic directions. For the first two however, the error reduction rate was observed to become less ideal for higher spline degrees. When including the Poisson matrix in the manufactured solution, we saw an earlier deviation from the mathematically predicted linear error decrease for higher spline degrees than for lower ones. All in all we view the Fourier-enhanced 3D solver presented in this paper as verified.

\section{Future Work}\label{sec:futurework}
The impending challenges regarding field solving for PICLS are threefold.\\
As already mentioned, the assumption of a Maxwellian for particle density in the initial Lagrangian needs to be relaxed in order to be strict on the full-f approach. This does not only affect the polarization equation of the solver but also the equations of motion in the particle pusher.\\
Furthermore, the Fourier approach to the solver, while saving a significant amount of computational cost without introducing inaccuracy, is not suitable to simulate scenarios that can not be modeled as periodic in at least one dimension or that are not well described by the assumption of axisymmetry in this dimension. A prominent example of this would be stellarators. To extend the applicability of PICLS to such problems, a full 3D solver option would need to be implemented. 
Finally, the simplification to electrostatic scenarios poses a significant limitation and will be lifted next. Like the delinearization of the polarization, this will require change in both the solver and the particle pusher, which will be the subject of future work.

\section{Acknowledgments}
This work has been carried out within the framework of the EUROfusion Consortium, partially funded by the European Union via the Euratom Research and Training Programme (Grant Agreement No 101052200 — EUROfusion). The Swiss contribution to this work has been funded by the Swiss State Secretariat for Education, Research and Innovation (SERI). Views and opinions expressed are however those of the author(s) only and do not necessarily reflect those of the European Union, the European Commission or SERI. Neither the European Union nor the European Commission nor SERI can be held responsible for them.

\appendix

\section{Appendix A: Values of $M^{(n)}$}
\label{sec:appendixA}

\subsection{Linear splines}
For the linear B-splines, ($p = 1$), we obtain
\begin{equation}
\begin{split}
   &m_0 = \frac{2}{3} , \quad m_1 = \frac{1}{6} \\
   &M^{(n)} = \frac{2\pi}{n_{\varphi}}\left(\cos\left(\frac{2 \pi}{n_{\varphi}}n\right) + 2\right)\frac{1}{3}
\end{split}
\end{equation}

\subsection{Quadratic splines}
For the quadratic ($p = 2$) B-splines:
\begin{equation}
\begin{split}
    &m_0 = \frac{11}{20}, \quad m_1 = \frac{13}{60}, \quad m_2 = \frac{1}{120} \\
    &M^{(n)} = \frac{2\pi}{n_{\varphi}}\left(\frac{1}{60} \cos\left(\frac{2 \pi}{n_{\varphi}}2n\right) + \frac{13}{30} \cos\left(\frac{2 \pi}{n_{\varphi}}n\right) + \frac{11}{20}\right)
\end{split}
\end{equation}
which is more conveniently rewritten by using:
\begin{equation}
\label{trig1}
\begin{split}
    \cos\left(\frac{2 \pi}{n_{\varphi}}2n\right)=2\cos^2\left(\frac{2 \pi}{n_{\varphi}}n\right)-1
\end{split}
\end{equation}
Leading to:
\begin{equation}
\begin{split}
    M^{(n)} = \frac{2\pi}{n_{\varphi}}\left(\frac{1}{30} \cos^2\left(\frac{2 \pi}{n_{\varphi}}n\right) + \frac{13}{30} \cos\left(\frac{2 \pi}{n_{\varphi}}n\right) + \frac{8}{15}\right)
\end{split}
\end{equation}

\subsection{Cubic splines}
For the cubic ($p = 3$) B-splines:
\begin{equation}
    m_0 = \frac{151}{315}, \quad m_1 = \frac{397}{1680}, \quad m_2 = \frac{1}{42}, \quad m_3 = \frac{1}{5040} 
\end{equation}
\begin{equation}
\begin{aligned}
    M^{(n)} =& \frac{2\pi}{n_{\varphi}}\left(\frac{1}{2520} \cos\left(\frac{2 \pi}{n_{\varphi}}3n\right) + \frac{1}{21} \cos\left(\frac{2 \pi}{n_{\varphi}}2n\right) \right. \\& \left. +\frac{397}{84} \cos\left(\frac{2 \pi}{n_{\varphi}}n\right) + \frac{151}{315}\right)
\end{aligned}
\end{equation}
which is more conveniently rewritten by using \cref{trig1} and
\begin{equation}
\label{trig2}
    \cos\left(\frac{2 \pi}{n_{\varphi}}3n\right)=4\cos^3\left(\frac{2 \pi}{n_{\varphi}}n\right)-3\cos\left(\frac{2 \pi}{n_{\varphi}}n\right)
\end{equation}
Leading to:
\begin{equation}
\begin{aligned}
    M^{(n)} =& \frac{2\pi}{n_{\varphi}}\left(\frac{1}{630} \cos^3\left(\frac{2 \pi}{n_{\varphi}}n\right)+\frac{2}{21} \cos^2\left(\frac{2 \pi}{n_{\varphi}}n\right)\right. \\&\left. +\frac{33}{70} \cos\left(\frac{2 \pi}{n_{\varphi}}n\right) + \frac{136}{315}\right)
\end{aligned}
\end{equation}

\section{Appendix B: Values of $D^{(n)}$}
\label{sec:appendixB}
\subsection{Linear splines}
For the linear B-splines, ($p = 1$), we obtain
\begin{equation}
\begin{split}
    &d_0 = 2,\quad  d_1 = -1 \\
    &D^{(n)} = \frac{n_{y}}{2\pi}\left(-2\cos\left(\frac{2 \pi}{n_{y}}n\right) + 2\right)
\end{split}
\end{equation}

\subsection{Quadratic splines}
For the quadratic ($p = 2$) B-splines:
\begin{equation}
\begin{split}
    &d_0 = 1, \quad d_1 = -\frac{1}{3}, \quad d_2 = -\frac{1}{6} \\
    &D^{(n)} = \frac{n_{y}}{2\pi}\left(-\frac{1}{3} \cos\left(\frac{2 \pi}{n_{y}}2n\right) - \frac{2}{3} \cos\left(\frac{2 \pi}{n_{y}}n\right) + 1\right)
\end{split}
\end{equation}
which is more conveniently rewritten by using \cref{trig1}, leading to:
\begin{equation}
\begin{split}
    D^{(n)} = \frac{n_{y}}{2\pi}\left(-2\cos^2\left(\frac{2 \pi}{n_{y}}n\right) - 2\cos\left(\frac{2 \pi}{n_{y}}n\right) + 4\right)\frac{1}{3}
\end{split}
\end{equation}

\subsection{Cubic splines}
For the cubic ($p = 3$) B-splines:
\begin{equation}
\begin{split}
    &d_0 = \frac{2}{3}, \quad d_1 = -\frac{1}{8}, \quad d_2 = -\frac{1}{5}, \quad d_3 = -\frac{1}{120}\\
    &D^{(n)} = \frac{n_{y}}{2\pi} ( -\frac{1}{60} \cos\left(\frac{2 \pi}{n_{y}}3n\right) - \frac{2}{5} \cos\left(\frac{2 \pi}{n_{y}}2n\right) -\\ 
    &\quad \frac{1}{4} \cos\left(\frac{2 \pi}{n_{y}}n\right) + \frac{2}{3} )
\end{split}
\end{equation}
which is more conveniently rewritten by using \cref{trig1} and
\begin{equation}
    \cos\left(\frac{2 \pi}{n_{y}}3n\right)=4\cos^3\left(\frac{2 \pi}{n_{y}}n\right)-3\cos\left(\frac{2 \pi}{n_{y}}n\right)
\end{equation}
Leading to:
\begin{equation}
\begin{split}
    &D^{(n)} = \frac{n_{y}}{2\pi} ( -\cos^3\left(\frac{2 \pi}{n_{y}}n\right)-12 \cos^2\left(\frac{2 \pi}{n_{y}}n\right) -\\
    &\quad 3\cos\left(\frac{2 \pi}{n_{y}}n\right) + 16\frac{1}{5} )
\end{split}
\end{equation}

 \bibliographystyle{elsarticle-num} 
 \bibliography{bib-file.bib}





\end{document}